\documentclass[twocolumn]{aastex631}
\usepackage{amsmath}
\usepackage{CJKutf8}

\shorttitle{AASTeX v6.3.1 Sample article}
\shortauthors{Wang et al.}

\graphicspath{{./}{figures/}}

\begin{document} 
\begin{CJK}{UTF8}{gbsn}
%\title{Milky Way Insights from Cepheids \uppercase\expandafter{\romannumeral1}: Infrared Period-Luminosity-Metallicity Relations and the evolution of metallicity gradient}

\title{Milky Way Classical Cepheids Distances from Calibrated Infrared Period-Luminosity-Metallicity Relations}

\author{Huajian Wang}
\affiliation{Purple Mountain Observatory, Chinese Academy of Sciences, Nanjing 210008, Peopleʼs Republic of China; hjwang@pmo.ac.cn}
\affiliation{University of Science and Technology of China, 96 Jinzhai Road, Hefei 230026, Peopleʼs Republic of China}

\author{Xiaodian Chen}
\affiliation{CAS Key Laboratory of Optical Astronomy, National Astronomical Observatories, Chinese Academy of Sciences, Beijing 100101, Peopleʼs Republic of China;}

\begin{abstract} 
Classical Cepheids (DCEPs) serve as fundamental standard candles for measuring cosmic distances and investigating the structure and evolution of the Milky Way disc. However, accurate distance estimation faces challenges due to severe extinction, particularly toward the Galactic center. Although the \textit{Gaia} Wesenheit magnitude reduces extinction effects, its reliance on a constant optical extinction law introduces significant uncertainties in regions of heavy obscuration. Infrared Period-Luminosity relations, combined with 3D extinction maps, offer an alternative, but these maps become unreliable beyond approximately 5 kpc. In this work, we calibrate the Period-Luminosity-Metallicity (PLZ) relations for DCEPs across three near-infrared bands ($J, H, K_S$) and four mid-infrared bands ($W1, W2, [3.6]$, and $[4.5]$). This includes the first calibration of the $W1$ and $W2$ bands. To correct for extinction, we employ the infrared multi-passband optimal distance method and the BP-RP method, which complement and validate each other. These homogeneous PLZ relations, combined with reliable extinction corrections, yield the most accurate Galactic DCEP distances to date, covering 3,452 DCEPs with an average relative distance error of 3.1\%. 
\end{abstract}

\section{Introduction} \label{sec:intro}
Classical Cepheids (DCEPs) are excellent distance indicators due to their well-known Period-Luminosity (PL) relation \citep{leavitt12}. Benefiting from the high-precision parallaxes of \textit{Gaia} \citep{Collaboration16}, recent studies have confirmed the influence of metallicity on the luminosity of DCEPs and have calibrated the Period-Luminosity-Metallicity (PLZ) relations across optical to infrared bands \citep[e.g.,][]{breuval22,ripepi22,bhar24,trentin24}. DCEPs play a crucial role in measuring the Hubble constant \citep{free2012,riess22} and serve as vital tracers for unveiling the structure and evolution of the Milky Way (MW) disc \citep[e.g.,][]{chen19,skowron19,lemasle22,drimmel24,huang24,zhou24}, where they are widely distributed. However, severe extinction in the MW disc, especially toward the Galactic center, significantly complicates distance estimations.

Currently, a widely adopted method for determining DCEP distances is based on the \textit{Gaia}'s Period-Wesenheit-Metallicity (PWZ) relation \citep[e.g.,][hereafter G23]{gaia23}, which reduces extinction effects by defining the Wesenheit magnitude. However, this method assumes a constant optical extinction law (i.e., $\lambda=1.9$), which is unreliable since $\lambda$ varies significantly across the MW disc \citep{wang17}. Another widely adopted method relies on infrared PL relations combined with 3D extinction maps \citep[hereafter S24]{skowron19,skowron24}. However, these maps are only reliable within approximately 5 kpc and cannot fully cover the entire MW disc \citep{zhang23}, leaving distance estimates from this method still debatable.

To improve distance accuracy with reliable extinction corrections, we adopt two complementary methods: the infrared multi-passband optimal distance method \citep[hereafter optimal distance method,][]{chen18,chen19} and the BP-RP method. The optimal distance method derives extinction by minimizing the standard deviation of distances derived from multi-passband infrared photometry and corresponding PLZ relations. The BP-RP method derives extinction from the color excess $E(G_{\text{BP}} - G_{\text{RP}})$ using the corresponding photometry and PLZ relations.

Although several studies have calibrated infrared PLZ relations \citep[e.g.,][]{breuval22,trentin24}, inconsistencies remain due to varying calibration methods, different residual parallax offset adoptions after parallax corrections \citep[hereafter L21]{lindegren21}, and discrepancies in the metallicity coefficients ($\gamma$). Moreover, the PLZ relations for the $W1$ and $W2$ bands from \textit{Wide-field Infrared Survey Explorer} have not been previously calibrated, limiting the utility of unWISE data \citep{schlafly19}, which benefits from co-added WISE images with significantly increased exposure time. Therefore, we systematically calibrate the PLZ relations across three near-infrared (NIR) bands ($J, H, K_S$) and four mid-infrared (MIR) bands ($W1, W2, [3.6]$, and $[4.5]$), incorporating open cluster Cepheids (OC-DCEPs) to leverage their statistically robust parallax measurements \citep{riess22,reyes23,wang24}. Using these seven homogeneous PLZ relations, combined with the optimal distance method and BP-RP method, we derive the most accurate Galactic DCEP distances to date.

This paper is organized as follows: In Section 2, we introduce the data used for calibration. In Section 3, we calibrate the PLZ relations for $J, H, K_S, W1, W2, [3.6]$, and $[4.5]$ bands. In Section 4, we apply these PLZ relations to derive distances for 3,452 Galactic DCEPs. In Section 5, we summarize this work.

\section{Data} \label{sec:Data}
This section describes the data we used to calibrate the infrared PLZ relations, including metallicity, extinction, parallax, and photometry.

\subsection{Metallicity and Extinction} \label{sec:metallicity}
The metallicities of 1002 DCEPs are compiled from \citet[hereafter T24a]{trentin24} and \citet[hereafter T24b]{trentin24b}. Among them, the metallicities of 726 DCEPs are obtained from  high-resolution (HiRes) spectroscopy \citep[e.g.,][]{greo18,trentin23,trentin24b}. The remaining metallicities are derived from the medium-resolution spectroscopy obtained with the \textit{Gaia} Radial Velocity Spectrometer \citep{recio23}. The Gaia metallicities are homogenized with the HiRes metallicities. For calibration purposes, only the 726 HiRes metallicities are used.

The extinctions are derived using the BP-RP method: First, the \textit{Gaia} BP and RP band PLZ relations of T24a are used to derive the intrinsic color. Then, the color excess is obtained by subtracting the intrinsic color from the observed color as: $E(G_{\text{BP}}-G_{\text{RP}}) = (m_{G_{\text{BP}}} - m_{G_{\text{RP}}})-(M_{G_{\text{BP}}} - M_{G_{\text{RP}}})$. $E(G_{\text{BP}}-G_{\text{RP}})$ is subsequently converted into extinction in various bands using the extinction coefficient $A_{\lambda}/E(G_{\text{BP}}-G_{\text{RP}})$ from \citet[see their Table 3]{wang19}. For the $m_{G_{\text{BP}}}$ and $m_{G_{\text{RP}}}$, we use intensity-averaged magnitudes \citep{Ripepi23} when available, otherwise, we use simple average magnitudes. We compare the extinctions obtained by this method with literature values \citep[e.g.,][]{greo18,ripepi21a}, and the results show excellent agreement.

\subsection{Parallax} \label{sec:Parallaxes}

For field DCEPs, the L21-corrected parallaxes from \textit{Gaia} data release 3 (DR3) are used, with the criterion RUWE \,\textless\, 1.4. However, the DCEPs used to calibrate the PLZ relations are very close (mostly less than 5 kpc) and young, meaning they are brighter than the optimal L21 corrections range (13 \,\textless\, $m_{G}$ \,\textless\, 17). This results in a residual parallax offset ($zp$) that needs to be subtracted \citep{lindegren21}. \cite{riess21} found $zp = -14\pm6\,\mu\textrm{as}$, a value supported by \cite{wang24}, who found $zp = -15\pm3\,\mu\textrm{as}$.

For OC-DCEPs, the L21-corrected parallaxes of their host OCs are used.
The OC's parallax is the average of its member stars, therefore, its statistical uncertainty benefits from $\sqrt{N}$, causing the error owing to the angular covariance of the \textit{Gaia} parallaxes to be dominant. Finally, the parallax errors of OC-DCEPs are around 10 $\mu$as. In addition, since the apparent magnitudes of most OC member stars fall within the optimal L21 corrections range, no additional $zp$ corrections are required. The release of DR3 has greatly helped the discovery of OC-DCEPs. To search OC-DCEPs, it is primarily necessary to compare the consistency of the five-dimensional parameters (ra, dec, $\mu_{{\alpha}^\ast}$, $\mu_\delta$, and $\varpi$) between OCs and DCEPs, as well as whether DCEPs are located on the instability strip of their host OC's CMD \citep{turner06}. \cite{reyes23} and \cite{wang24} found 34 and 43 OC-DCEPs, respectively. Since there is some overlap between these two works, we finally compile a total of 52 OC-DCEPs.

\subsection{Photometry} \label{sec:Photometry}
\textit{NIR J, H and $K_S$}. Similar to T24a, we primarily obtain $J$, $H$ and $K_S$ photometry from the literature \citep{van07,GC17,greo18,ripepi21a}. The photometry of remaining DCEPs is obtained by 1” matching with the 2MASS All-Sky Point Source Catalog \citep{skru06}.

\textit{WISE MIR W1 and W2}. We obtain the photometry of $W1$ and $W2$ from the AllWISE Source Catalog \citep{cutri13} with a matching radius of 1''. In addition, for sources brighter than $W1$ = 8 mag and $W2$ = 7 mag and were observed during the Post-Cryo mission phase (89.4$^{\circ}$ \,\textless\, $\lambda$ \,\textless\, 221.8$^{\circ}$ and 280.6$^{\circ}$ \,\textless\, $\lambda$ \,\textless\, 48.1$^{\circ}$), we use data from the WISE All-Sky Release Source Catalog to avoid the unreliable AllWISE measurements\footnote{\url{https://wise2.ipac.caltech.edu/docs/release/allwise/expsup/sec2_2.html}}. Although the unWISE Catalog \citep{schlafly19} is still being updated and has more than five times the exposure time of the AllWISE Catalog, the unWISE Catalog is mostly utilized for fainter magnitudes, whereas the AllWISE Catalog is more reliable at brighter magnitudes. We do not use unWISE photometry for calibration because the DCEPs for calibration are bright.

\textit{Spitzer MIR [3.6] and [4.5]}. We obtain the photometry of $[3.6]$ and $[4.5]$ from Spitzer's GLIMPSE (Galactic Legacy Infrared Midplane Survey Extraordinaire) program. We matched the following GLIMPSE subprojects with a matching radius of 1'': GLIMPSE I, GLIMPSE II, GLIMPSE 3D, GLIMPSE360, Vela-Carina, Deep GLIMPSE, SMOG, Cygnus-X, GLIMPSE Proper, and APOGLIMPSE \citep{benjamin03,churchwell09}. In addition, we add the $[3.6]$ and $[4.5]$ photometry of 37 DCEPs from \cite{monson12}.

\section{PLZ relations}\label{sec:PLZ relation}

\subsection{Method}\label{sec:method}
We adopt the same method for calibrating the PLZ relations as described in \cite{wang24}, referring to the methods in \cite{riess22} and \cite{ripepi22}. The method is described as follows. First, we define the photometric parallax as:  
\begin{equation}
\varpi_{\textrm{phot}} = 10^{-0.2(m_\lambda-A_\lambda - M_\lambda)+2},
\end{equation} 
where $A_\lambda$ is the extinction and $m_\lambda$ is the apparent magnitude. The absolute magnitude, $M_\lambda$, can be defined as
\begin{equation}
M_\lambda=\alpha (\log P_{\textrm{F}} - 1) + \beta + \gamma[\textrm{Fe/H}],
\end{equation}
where $P_{\textrm{F}}$ represents the period of the fundamental mode (F-mode). Using the \texttt{optimize.minimize} method from the \texttt{Python Scipy} library, we minimize the following quantity:
\begin{equation}
\chi^2 =
\begin{cases}
\sum{\frac{\left(\varpi_{\textrm{OC}}-\varpi_{\textrm{phot}}\right)^2}{{\sigma_{\varpi_{\textrm{OC}}}^2 + \sigma_{\varpi_{\textrm{phot}}}^2 }}} & \textrm{for OC-DCEPs},\\
\sum{\frac{\left(\varpi_{\textrm{DCEP}}-\varpi_{\textrm{phot}}+zp\right)^2}{{\sigma_{\varpi_{\textrm{DCEP}}}^2 + \sigma_{\varpi_{\textrm{phot}}}^2 }}} & \textrm{for field DCEPs},
\end{cases}
\end{equation}
where $zp$ is the residual parallax offset in field DCEPs and we adopt $zp = -15\pm3\,\mu\textrm{as}$. $\sigma_{\varpi_{\textrm{phot}}}$ can be defined based on the error propagation formula as: $\sigma_{\varpi_{\textrm{phot}}} = 0.46 \times \sqrt{\sigma_{m_\lambda}^2 + \sigma_{A_\lambda}^2 + \sigma_{M_\lambda}^2} \times \varpi_{\textrm{phot}}$, where $\sigma_{m_\lambda}$ represents the error of apparent magnitude. The extinction error, $\sigma_{A_\lambda}$, is obtained by propagating the errors of $E(G_{\text{BP}}-G_{\text{RP}})$ and $A_{\lambda}/E(G_{\text{BP}}-G_{\text{RP}})$. We conservatively set the $E(G_{\text{BP}}-G_{\text{RP}})$ error to 0.1 mag. The error in absolute magnitude, $\sigma_{M_\lambda}$ (width), is the intrinsic dispersion resulting from the finite width of the instability strip of the DCEPs. We adopt widths of 0.11, 0.09, and 0.07 mag \citep{persson04} for the $J$, $H$, and $K_S$ bands; a width of 0.07 mag \citep{scow2011,monson12} for the $[3.6]$ and $[4.5]$ bands; and widths of 0.08 and 0.11 mag \citep{wang18} for the $W1$ and $W2$ bands.

\begin{figure*}[htb]
\figurenum{1}
\gridline{\fig{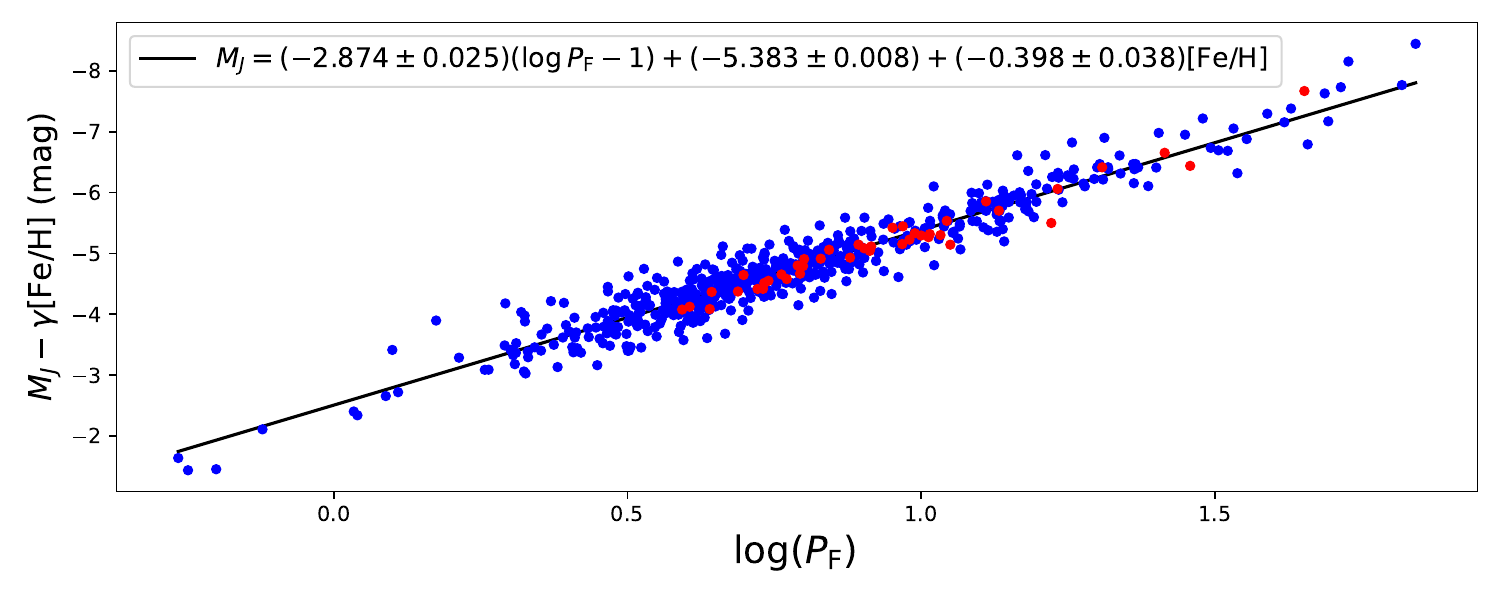}{0.5\textwidth}{} \hfill \fig{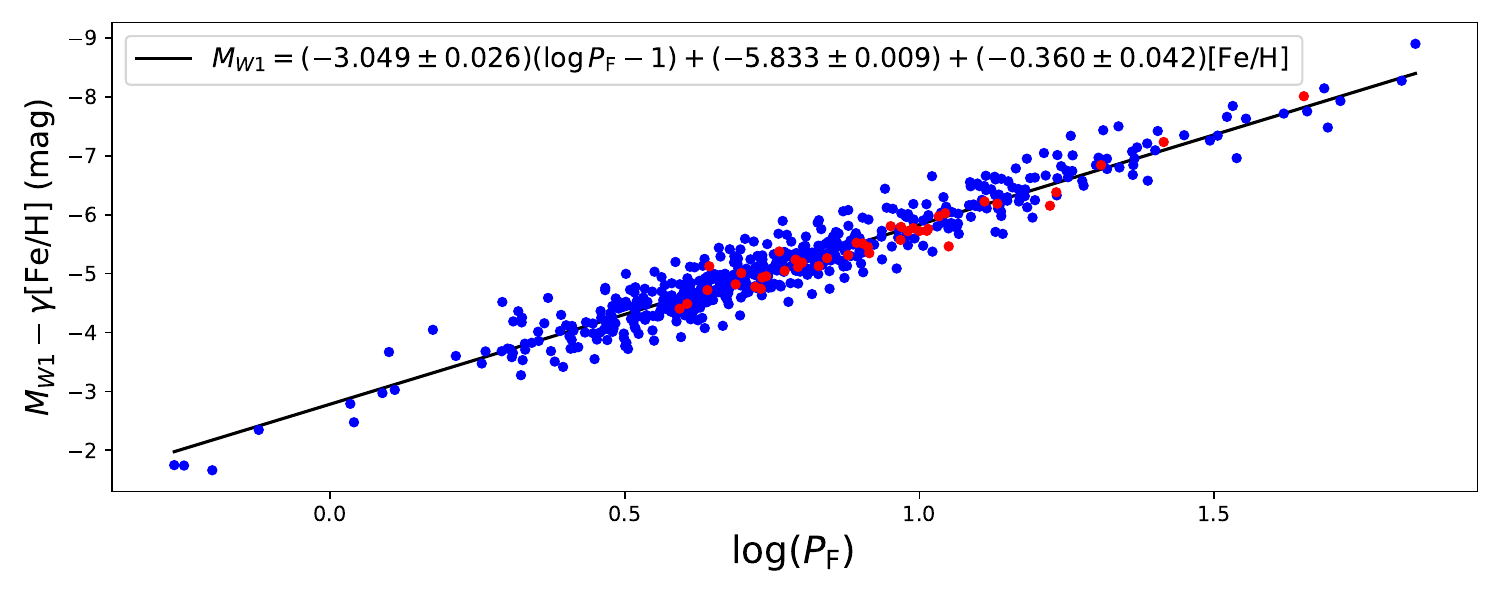}{0.5\textwidth}{}}
\vspace{-1cm} % 调整上下图之间的间隙
\gridline{\fig{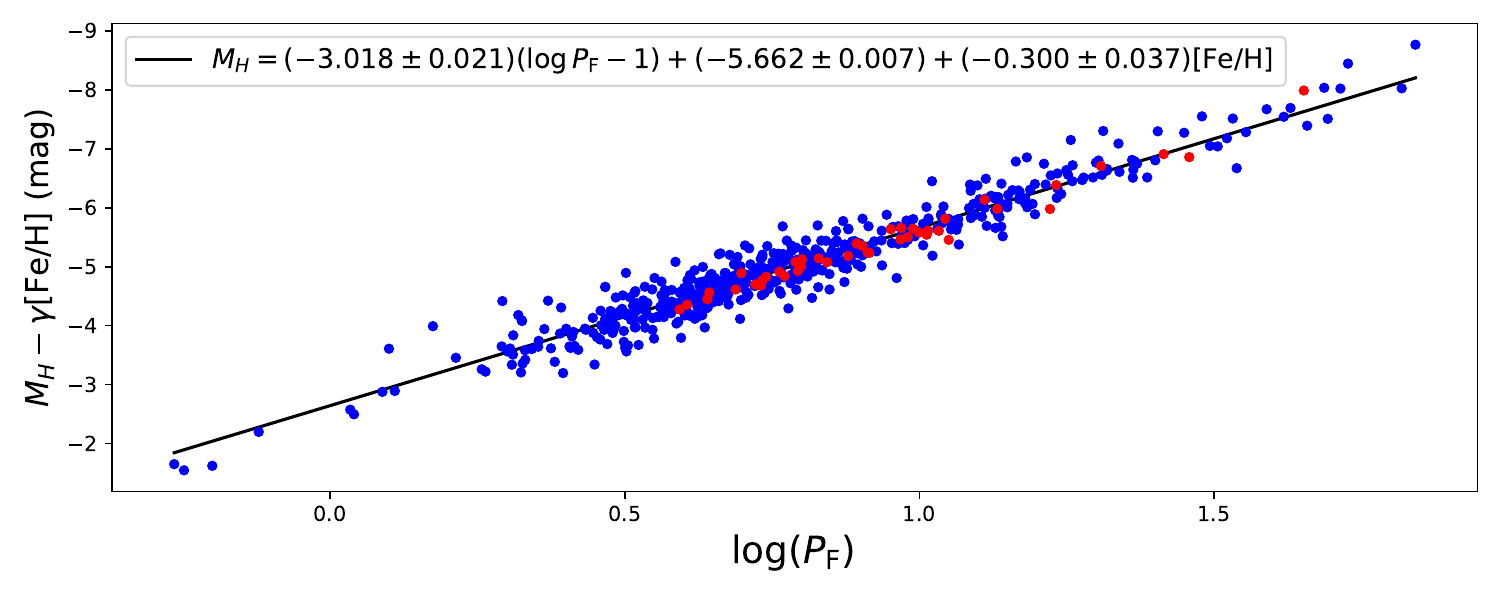}{0.5\textwidth}{} \hfill \fig{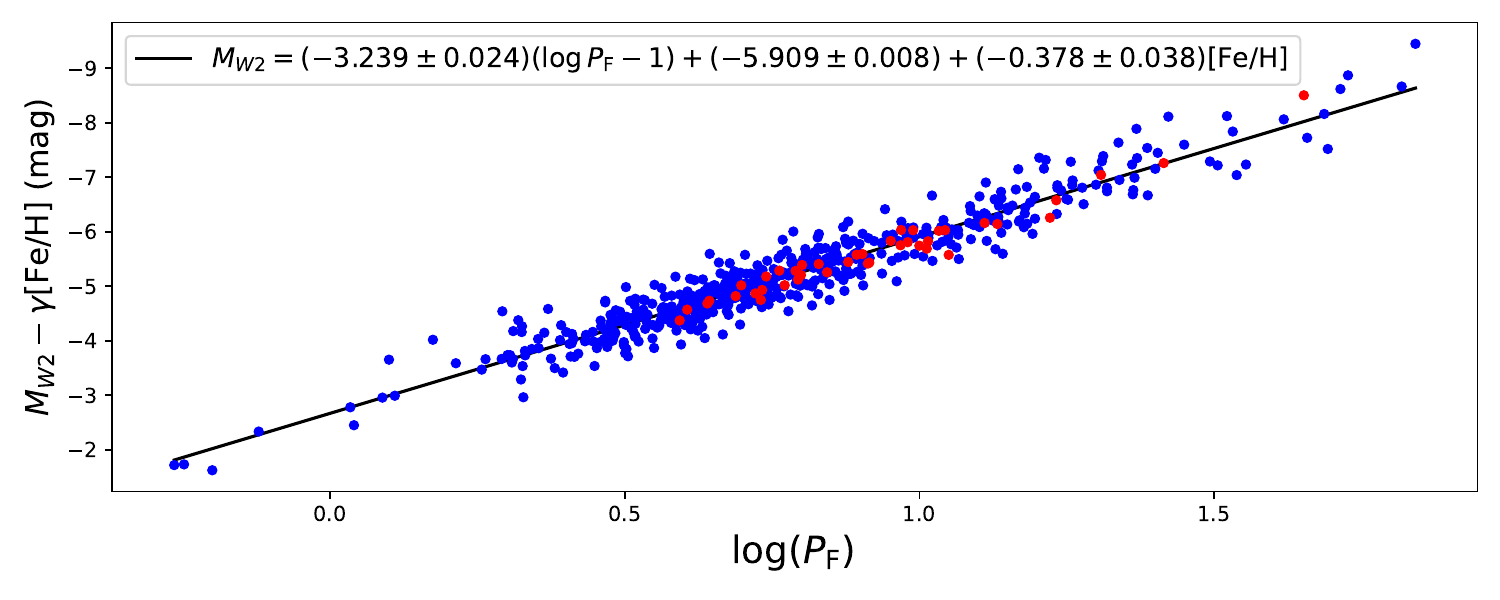}{0.5\textwidth}{}}
\vspace{-1cm} % 调整上下图之间的间隙
\gridline{\fig{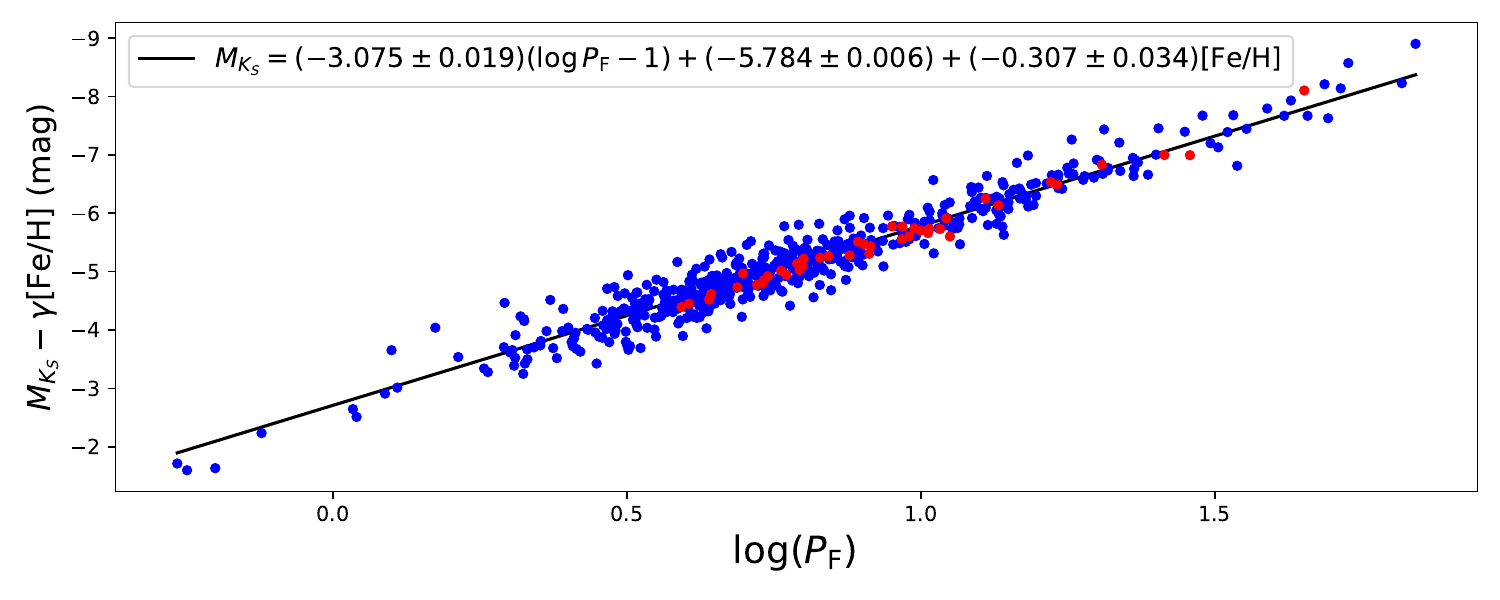}{0.5\textwidth}{} \hfill \fig{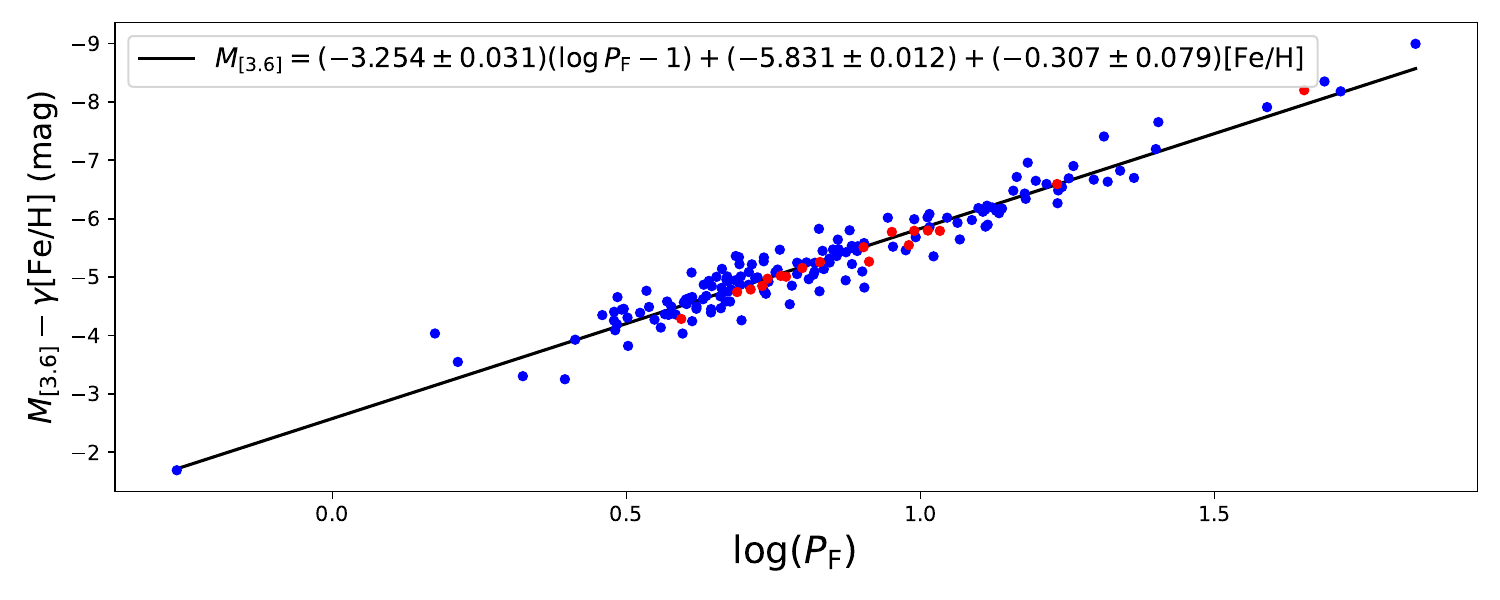}{0.5\textwidth}{}}
\vspace{-1cm} % 调整上下图之间的间隙
\gridline{\fig{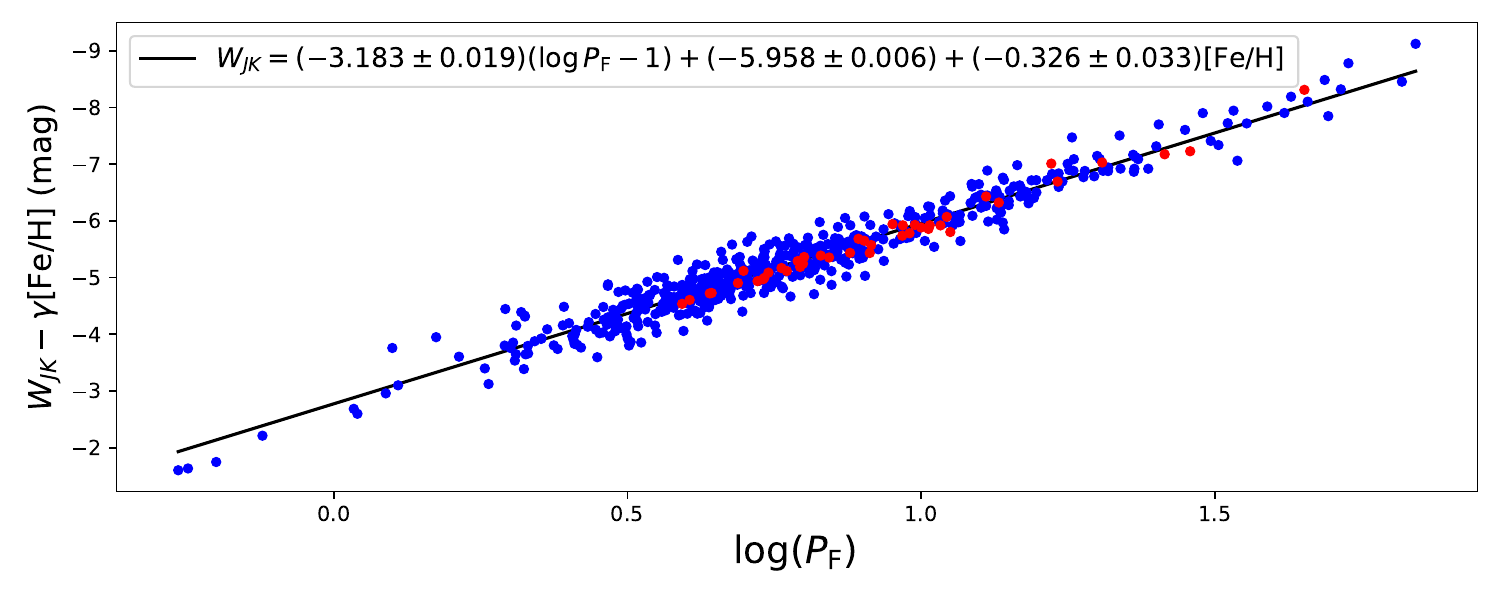}{0.5\textwidth}{} \hfill \fig{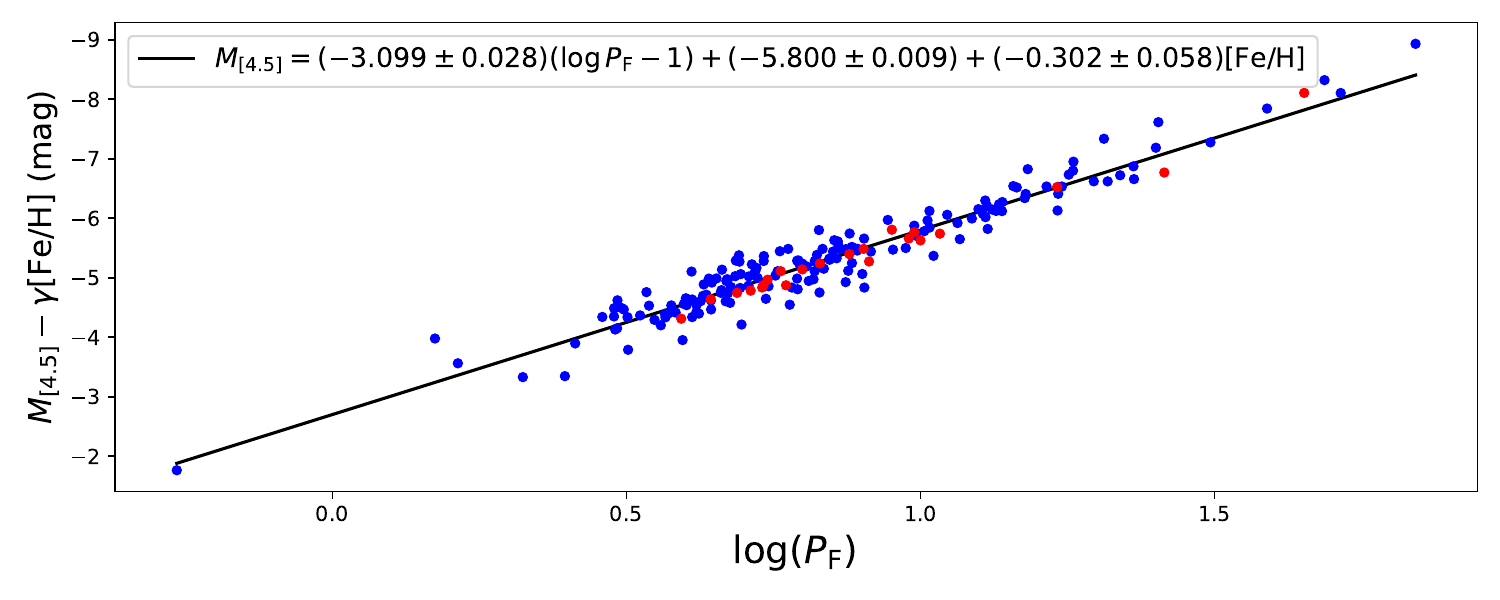}{0.5\textwidth}{}}
\caption{Calibration of infrared PLZ relations. The ordinate is the absolute magnitude minus the contribution of metallicity. Blue dots represent field DCEPs, red dots represent OC-DCEPs. Note that we also plot the 1O-mode DCEPs with $\log P_{\textrm{F}} \lesssim 0.3$, but they are not used in the fitting.
\label{fig:1}}
\end{figure*}

For multiple mode DCEPs (F1O, 1O2O), we use their longer periods. Then we use the equation $P_{\textrm{F}} = P_{\textrm{1O}}/(0.716-0.027\log P_{\textrm{1O}})$ \citep{feast97} to fundamentalize the 1O-mode periods. 
In our calibration sample, the period range of F-mode DCEPs is $0.3 \lesssim \log P_{\textrm{F}} \lesssim 1.7$, and the period range of 1O-mode DCEPs is $-0.3 \lesssim \log P_{\textrm{F}} \lesssim 1.0$ (mainly distributed in $0.3 \lesssim \log P_{\textrm{F}} \lesssim 1.0$). \cite{ripepi22a} found that the PL relation for 1O-mode DCEPs in the Large Magellanic Cloud (LMC) exhibits a break at $P_{1O}$ = 0.58 (i.e., $\log P_{\textrm{F}}$ $\approx$ -0.10). Therefore, although the break has only been observed in the Magellanic Cloud Cepheids \citep{ngeow15,bhar16,ripepi22a}, we only add 1O-mode DCEPs with periods in the range $0.3 \lesssim \log P_{\textrm{F}} \lesssim 1.0$, which can retain most of the 1O-mode DCEPs and prevent the potential confusion caused by the possible break in 1O-mode DCEPs at shorter periods. To remove possible outlier measurements, we apply the 3-sigma clipping method for iterations. After several iterations, the number of samples converged. Figure \ref{fig:1} shows our PLZ relations \footnote{In addition, we also calibrate a period–Wesenheit–metallicity (PWZ) relation of $W_{JK}$. The apparent Wesenheit magnitude $W_{JK} = m_{K_S} - \lambda \times(m_J - m_{K_S})$, is the reddening-free magnitude \citep{madore82,majaess08}. $\lambda = A_{K_S}/E(J - K_S)$ and we adopt $\lambda = 0.473$ derived from \cite{wang19}. We adopt a width of 0.086 mag \citep{breuval22} for the $W_{JK}$}, whereas Table \ref{tab:1} presents the fitting results (e.g., $\alpha$, $\beta$, and $\gamma$).

\begin{deluxetable*}{cccccccc}[t]
\setlength{\tabcolsep}{6pt}
\tablenum{1}
\tablecaption{Infrared PLZ relations fitting parameters \label{tab:1}}
\tablewidth{0pt}
\tabletypesize{\normalsize}
\tablehead{
\colhead{Band} & \colhead{$\alpha$}& \colhead{$\beta$}& \colhead{$\gamma$}& \colhead{Mode}& \colhead{$\sigma$}& \colhead{$\mu_{\textrm{LMC}}$}& \colhead{$N$}\\
\colhead{} &\colhead{} &\colhead{(mag)} &\colhead{}&\colhead{} &\colhead{(mag)} &\colhead{(mag)} &\colhead{}}
\startdata
$J$ & $-2.913  \,\pm\,  0.028$ & $-5.365  \,\pm\,  0.009$ & $-0.448  \,\pm\,  0.051$ & F & 0.240 & 18.458  $\,\pm\,$  0.027 & 382 DCEPs + 36 OC-DCEPs \\ 
& $-2.874 \,\pm\,  0.025$ & $-5.383  \,\pm\,  0.008$ & $-0.398  \,\pm\,  0.038$ & F+1O & 0.256 & 18.489  $\,\pm\,$  0.023 &  536 DCEPs + 42 OC-DCEPs \\ \hline
$H$ & $-3.073  \,\pm\,  0.023$ & $-5.647  \,\pm\,  0.008$ & $-0.345  \,\pm\,  0.043$ & F & 0.228 & 18.414  $\,\pm\,$  0.022 & 379 DCEPs + 36 OC-DCEPs \\ 
& $-3.018 \,\pm\,  0.021$ & $-5.662  \,\pm\,  0.007$ & $-0.300  \,\pm\,  0.037$ & F+1O & 0.248 & 18.458  $\,\pm\,$  0.019 &  535 DCEPs + 42 OC-DCEPs \\ \hline
$K_S$& $-3.102  \,\pm\,  0.022$ & $-5.770  \,\pm\,  0.007$ & $-0.353  \,\pm\,  0.044$ & F & 0.236 & 18.463  $\,\pm\,$  0.021 & 379 DCEPs + 36 OC-DCEPs \\ 
& $-3.075 \,\pm\,  0.019$ & $-5.784  \,\pm\,  0.006$ & $-0.307  \,\pm\,  0.034$ & F+1O & 0.248 & 18.479  $\,\pm\,$  0.018 &  532 DCEPs + 42 OC-DCEPs \\ \hline
$W1$& $-3.074  \,\pm\,  0.028$ & $-5.813  \,\pm\,  0.009$ & $-0.445  \,\pm\,  0.052$ & F & 0.251 & 18.408  $\,\pm\,$  0.026 & 375 DCEPs + 35 OC-DCEPs \\ 
& $-3.049  \,\pm\,  0.026$ & $-5.833  \,\pm\,  0.009$ & $-0.360  \,\pm\,  0.042$ & F+1O & 0.263 & 18.464  $\,\pm\,$  0.020 &  530 DCEPs + 41 OC-DCEPs \\ \hline
$W2$& $-3.276  \,\pm\,  0.026$ & $-5.894  \,\pm\,  0.008$ & $-0.459  \,\pm\,  0.048$ & F & 0.293 & 18.391  $\,\pm\,$  0.026 & 383 DCEPs + 35 OC-DCEPs \\ 
& $-3.239  \,\pm\,  0.024$ & $-5.909  \,\pm\,  0.008$ & $-0.378  \,\pm\,  0.038$ & F+1O & 0.289 & 18.460  $\,\pm\,$  0.020 &  536 DCEPs + 41 OC-DCEPs \\ \hline
$[3.6]$& $-3.275  \,\pm\,  0.034$ & $-5.819  \,\pm\,  0.012$ & $-0.322  \,\pm\,  0.078$ & F & 0.233 & 18.386  $\,\pm\,$  0.036 & 115 DCEPs + 18 OC-DCEPs \\ 
& $-3.254  \,\pm\,  0.031$ & $-5.831  \,\pm\,  0.012$ & $-0.307  \,\pm\,  0.079$ & F+1O & 0.246 & 18.395  $\,\pm\,$  0.034 &  153 DCEPs + 18 OC-DCEPs \\ \hline
$[4.5]$& $-3.127  \,\pm\,  0.032$ & $-5.786  \,\pm\,  0.011$ & $-0.358  \,\pm\,  0.068$ & F & 0.232 & 18.381  $\,\pm\,$  0.032 & 126 DCEPs + 20 OC-DCEPs \\
& $-3.099  \,\pm\,  0.028$ & $-5.800  \,\pm\,  0.009$ & $-0.302  \,\pm\,  0.058$ & F+1O & 0.241 & 18.415  $\,\pm\,$  0.030 &  167 DCEPs + 22 OC-DCEPs \\ \hline
$W_{JK}$& $-3.241  \,\pm\,  0.022$ & $-5.947  \,\pm\,  0.007$ & $-0.362  \,\pm\,  0.045$ & F & 0.220 & 18.414  $\,\pm\,$  0.020 & 370 DCEPs + 36 OC-DCEPs \\ 
& $-3.183  \,\pm\,  0.019$ & $-5.958  \,\pm\,  0.006$ & $-0.326  \,\pm\,  0.033$ & F+1O & 0.236 & 18.431  $\,\pm\,$  0.017 &  526 DCEPs + 42 OC-DCEPs \\
\enddata
\tablecomments{$\alpha$, $\beta$, $\gamma$, and $\sigma$ are the slope, intercept, metallicity coefficient, and standard deviation, respectively. $\mu_{\textrm{LMC}}$ is the distance modulus of the LMC derived from our calibrated PLZ relations. $N$ is the number of DCEPs and OC-DCEPs used in the calibration.}
\end{deluxetable*} 

\subsection{Reliability Testing of PLZ relations on LMC DCEPs}\label{sec:LMC}
 
The photometry of LMC DCEPs \citep[][$N \sim 4600$]{udalski18} are obtained by 1'' matching with the 2MASS All-Sky Point Source Catalog \footnote{\url{https://irsa.ipac.caltech.edu/data/2MASS/docs/releases/allsky/doc/sec2_2.html}} ($J, H, K_S$), unWISE Catalog \footnote{\url{https://irsa.ipac.caltech.edu/data/WISE/unWISE/overview.html}. Since the apparent magnitudes of LMC DCEPs are all faint (mainly distributed in $12<W1<16$), we use the unWISE catalog, which performs better in faint magnitudes. Following the advice of the unWISE team \citep{unwise}, we subtract 0.004 and 0.032 mag from $W1$ and $W2$, respectively, to eliminate the systematic difference between unWISE and ALLWISE.} ($W1$ and $W2$), and SAGE IRAC Single Frame + Mosaic Photometry Catalog \footnote{\url{https://irsa.ipac.caltech.edu/data/SPITZER/SAGE/}} ($[3.6]$ and $[4.5]$). The dereddened magnitude ($m_\lambda - A_\lambda$) is obtained using the extinction map from \cite{skowron21}. Assuming a mean metallicity of $[\textrm{Fe/H]}_{\textrm{LMC}} = -0.409 \pm 0.003$ dex \citep{roman22}, we derive the absolute magnitudes ($M_\lambda$) using our PLZ relations. The distance modulus for each DCEP is estimated as ($m_\lambda-A_\lambda)-M_\lambda$. The median distance modulus is adopted as our estimate of $\mu_{\textrm{LMC}}$, with error determined through 10,000 Monte Carlo runs. The results, listed in \ref{tab:1}, are consistent with the geometric distance modulus from \cite{pie19}, confirming the reliability of our PLZ relations.

\subsection{The influence of $zp$ adoption and parallax cut}\label{sec:zp}
In this work, we adopt $zp = -15\,\mu\textrm{as}$ from \cite{wang24}, which is similar to the commonly used $zp = -14\,\mu\textrm{as}$ \citep{riess21,breuval22,ripepi22}. We also test another common value, $zp = -22\,\mu\textrm{as}$, which results in a decrease in $\gamma$ (in an absolute sense) and an increase in the measured $\mu_{\textrm{LMC}}$ by approximately 0.06 mag. However, the results still remain within 3$\sigma$ of the \cite{pie19} estimation. Conversely, if we adopt $zp = 0\,\mu\textrm{as}$, $\gamma$ increases (in an absolute sense), and $\mu_{\textrm{LMC}}$ decreases by approximately 0.11 mag, resulting in a discrepancy greater than 3$\sigma$ compared to the \cite{pie19} estimation. The exact value of $zp$ depends on the sample selection. Since $zp = -15\,\mu\textrm{as}$ was derived based on the T24a sample, which is also the primary source for calibration in this work, we adopt $zp = -15\,\mu\textrm{as}$ as the most appropriate choice.

We choose not to apply any parallax cuts, such as $\varpi/\sigma_{\varpi} > 10$, because such a criterion would restrict the sample to close to the Sun, significantly reducing the metallicity range of the sample. This, in turn, limits our ability to estimate $\gamma$. If we were to adopt $\varpi/\sigma_{\varpi} > 10$, we would obtain a smaller $\gamma$ (in an absolute sense).

\subsection{Comparison with the literature $\gamma$ in the infrared bands}\label{sec:com}
\begin{figure}[htb]
\figurenum{2}
\gridline{\fig{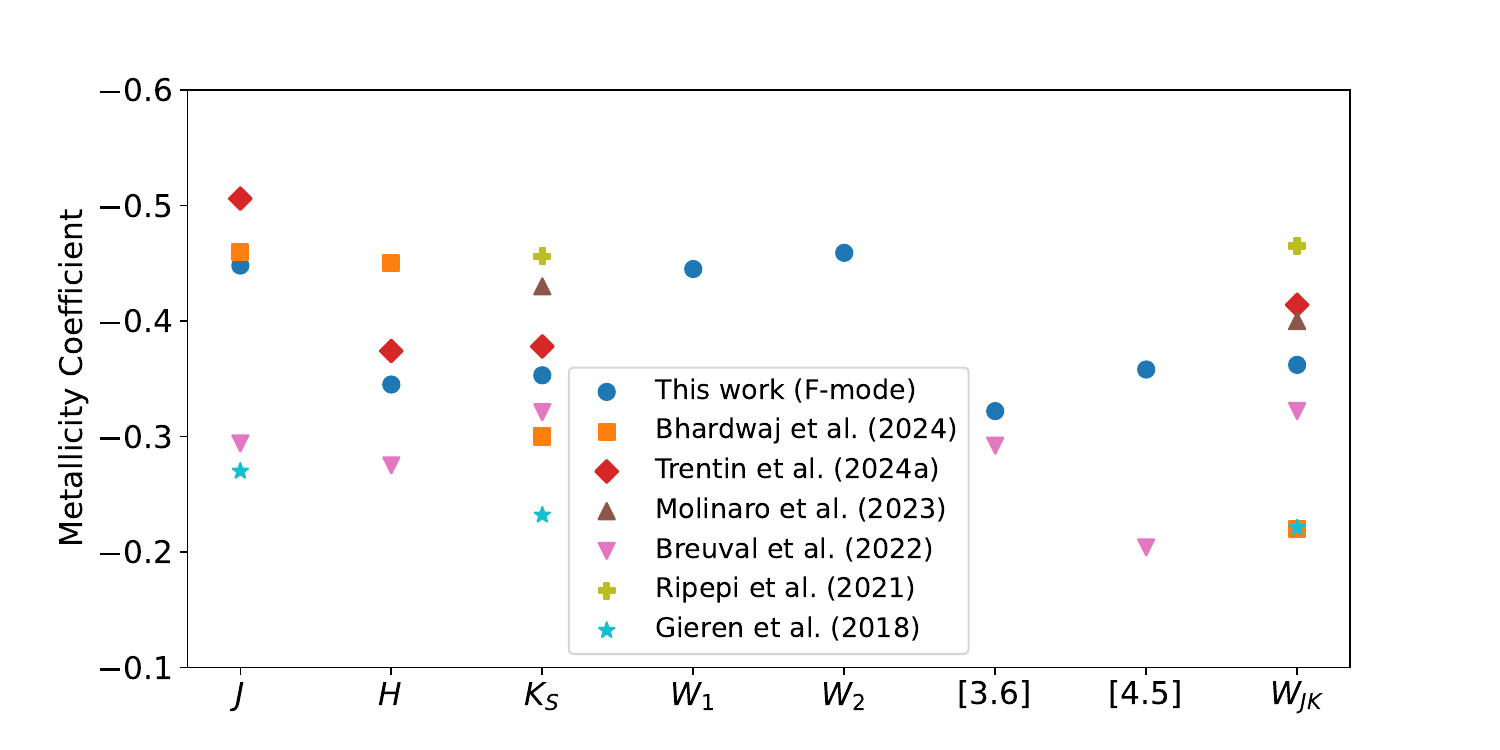}{0.5\textwidth}{}}
\caption{This figure compares the metallicity coefficients in infrared bands from recent studies  \citep{gieren18,ripepi21a,breuval22,molinaro23,trentin24,bhar24}.
\label{fig:2}}
\end{figure}

Figure \ref{fig:2} compares the infrared metallicity coefficients obtained in this work with those from other recent studies. \cite{gieren18} and \cite{breuval22} investigated the metallicity dependence of the PL/PW relations by comparing the metallicity of the MW, LMC, and SMC. Their fitted $\gamma$ in the infrared bands ($J$, $H$, $K_S$, $[3.6]$, $[4.5]$ and $W_{JK}$) range from -0.20 to -0.32, which are smaller (in an absolute sense) than our results (-0.30 to -0.46). Besides, the $\gamma$ obtained by the C-MetaLL project \citep[Cepheids–metallicity in the Leavitt Law,][]{ripepi21a,molinaro23,trentin24,bhar24} in the infrared bands ($J$, $H$, $K_S$ and $W_{JK}$) are slightly larger (in an absolute sense) than ours, for example, the $\gamma$ obtained by T24a range from -0.37 to -0.51. As discussed in the preceding subsection, a larger $zp$ reduces $\gamma$ (in an absolute sense), indicating that the precision of the parallax has a significant impact on the fitting results of $\gamma$. In the future, using more accurate parallax can help us better understand the metallicity dependence of the PL relations.

\section{Distance of Galactic DCEPs} \label{sec:distance}

In this section, we use our F+1O-mode PLZ relations to derive the distances of Galactic DCEPs. A total of 3,645 Galactic DCEPs, all confirmed in the optical range, are adopted from \cite{pie21}. The photometry of these DCEPs is obtained following a process similar to that described in Section \ref{sec:Photometry}, with specific adjustments for the $W1$ and $W2$ bands: unWISE photometry is adopted for $W1$ if $W1 > 8.25$ mag, and for $W2$ if $W2 > 8.0$ mag, as it provides better-quality photometry in crowded regions \citep{schlafly19}. Notably, the metallicities of 979 DCEPs are available from T24a and T24b, while the metallicities of the remaining DCEPs are estimated using the MW disc metallicity gradient from T24b: $[\textrm{Fe/H}] = (-0.064 \pm 0.003)R_{GC} + (0.530 \pm 0.029)$. Two methods, the optimal distance method and the BP-RP method, are used to derive the extinction and distance.

\subsection{Methods for distances derivation}\label{dis-method}
\textit{Optimal Distance Method}. The distance and distance error (in kpc) are defined as:
\begin{equation}
	\begin{split}
	d_\lambda &= 10^{0.2(m_\lambda - A_\lambda - M_\lambda) - 2},\\
	\sigma_{d_\lambda} &= 0.46 \times \sqrt{\sigma_{m_\lambda}^2 + \sigma_{A_\lambda}^2 + \sigma_{M_\lambda}^2} \times d_\lambda,
	\end{split}
\end{equation}
where $\lambda$ represents the $J, H, K_S, W1, W2, [3.6]$, and $[4.5]$ bands. In this equation, $M_\lambda$ is the absolute magnitude derived from our PLZ relations, and $A_\lambda$ is the extinction derived using the extinction law from \citet{wang19} and $A_K$. The values of $A_K$ and the weighted average distance are derived by minimizing the standard deviation of $d_\lambda$ across the seven infrared bands. It should be noted that this method requires at least one NIR band, as the flat extinction curve in the MIR makes MIR-only fitting unreliable. Thus, DCEPs with only MIR photometry are excluded from this method. Furthermore, the quality of the fit is also affected by the quality of the NIR photometry. High-quality NIR photometry is defined as measurements with \text{ph\_qual} of A, B, C, or D and \text{cc\_flg} = 0, indicating that the measurement in this band is valid and free from contamination or artifacts (or undetected). Therefore, in the NIR bands, we prioritize high-quality photometry for fitting. However, if high-quality NIR photometry is limited (e.g., only one high-quality NIR measurement among the seven infrared bands) or unavailable, we include all available NIR photometry in the fitting process. Using this method, we derived the extinctions and distances for 3,539 DCEPs.

\textit{BP-RP Method}. This method is described in Section \ref{sec:metallicity}. Here, in the NIR bands, we only use high-quality NIR photometry to derive distance. Notably, since the metallicity coefficients in the BP and RP bands are similar, the impact of metallicity on $E(G_{\text{BP}}-G_{\text{RP}})$ is negligible. In addition, it should be noted that $m_{G_{\text{BP}}}$ tends to become systematically brighter at the faint end \citep[$m_{G_{\text{BP}}} > 20.3$ mag, see Section 8.1 of][]{photometry}, which causes the BP-RP method to be unreliable for faint sources. Therefore, we limit our sample to DCEPs with $m_{G_{\text{BP}}} < 20.3$ mag. Using this method, we derived extinctions and distances for 3,275 DCEPs.

\begin{figure}[htb]
\figurenum{3}
\gridline{\fig{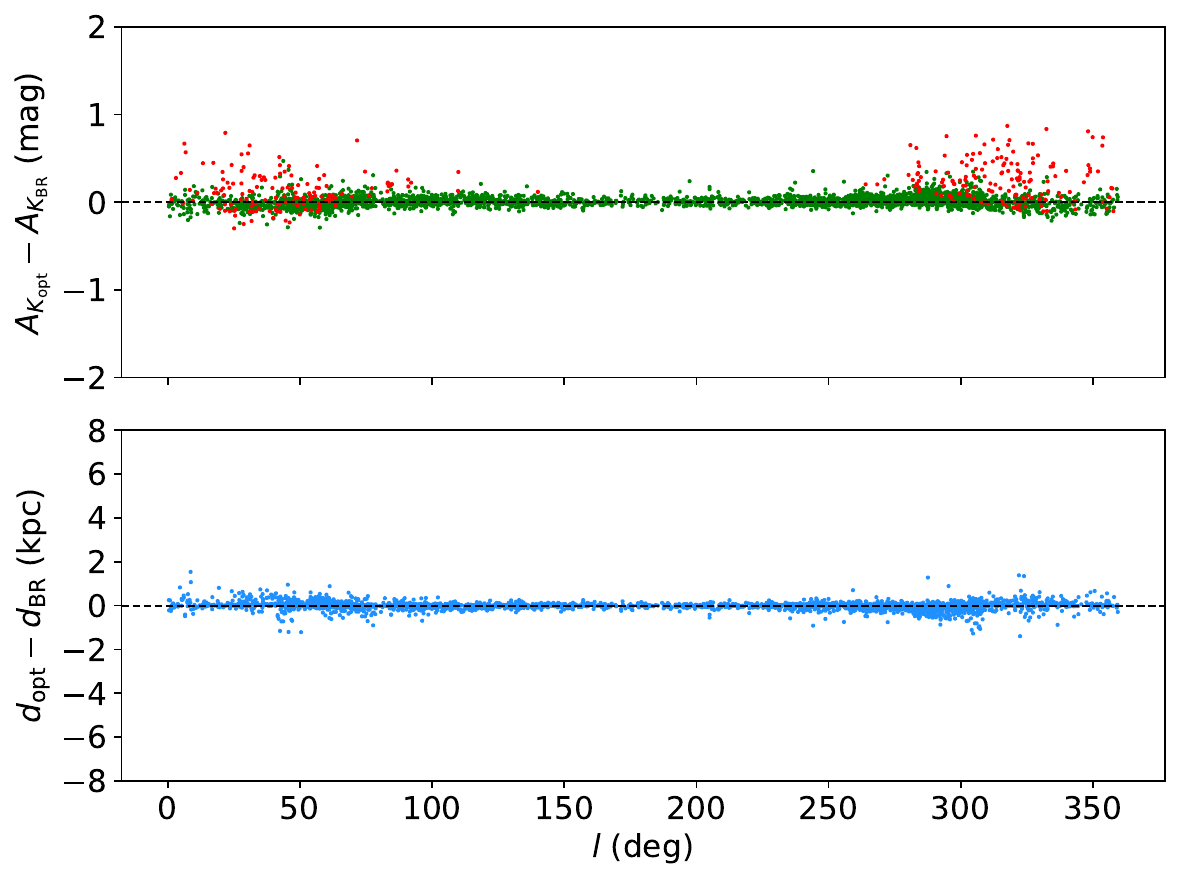}{0.45\textwidth}{}}
\caption{The top panel compares extinctions derived by the optimal distance method with those derived by the BP-RP method. Green dots represent 3,045 DCEPs that meet the criterion $m_{G_{\text{BP}}} < 20.3$ mag, showing an average difference of $0.010\pm0.062$ mag. In contrast, red dots represent 309 DCEPs with $m_{G_{\text{BP}}} > 20.3$ mag. The bottom panel illustrates a comparison of the distances for these 3,045 DCEPs, with an average difference of $-16\pm191$ pc.
\label{fig:3}}
\end{figure}

\begin{figure}[htb]
\figurenum{4}
\gridline{\fig{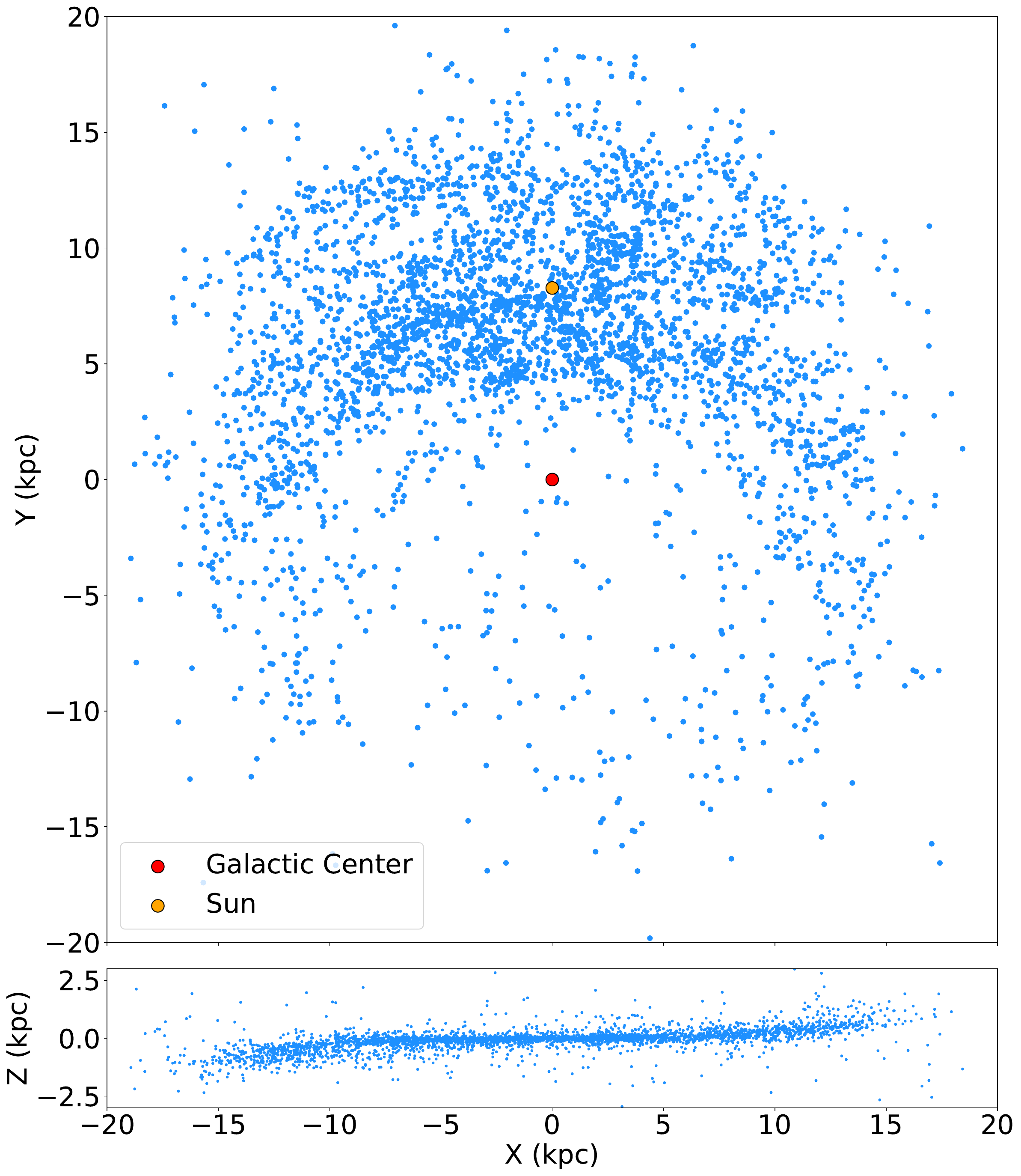}{0.45\textwidth}{}}
\caption{The top panel presents a bird's-eye view of the spatial distribution of 3,452 DCEPs across the MW, with the Sun located at Y = 8.277 kpc \citep{gravity}. The bottom panel illustrates the MW warp as traced by these DCEPs.
\label{fig:4}}
\end{figure}

\subsection{Sample Selection and Final Distances}\label{final distances}
We retain only DCEPs with a relative distance error below 10\% ($\sigma_d/d < 0.1$), where the distance error $\sigma_d$ is defined as the larger value between the standard deviation and the weighted error ($\sqrt{1 / \sum (1 / \sigma_{d_{\lambda}}^2)}$). Additionally, we limit our sample to DCEPs within 25 kpc of the Galactic center. As a result, 3,434 and 3,063 DCEPs are retained for the optimal distance and BP-RP methods, respectively. Among these, 3,045 DCEPs have distances derived from both methods. The top panel of Figure \ref{fig:3} compares the extinctions obtained from the optimal distance method ($A_{K_{opt}}$) and the BP-RP method ($A_{K_{BR}}$). Red dots represent DCEPs with $m_{G_{\text{BP}}} > 20.3$ mag, primarily located towards the Galactic center. These red dots with larger extinction differences show significantly smaller $A_{K_{BR}}$ compared to $A_{K_{opt}}$, indicating that the systematic brightening of 
$m_{G_{\text{BP}}}$ at the faint end reduces $A_{K_{BR}}$. For green dots ($m_{G_{\text{BP}}} < 20.3$ mag), the extinctions derived from both methods are in excellent agreement, with an average difference of $0.010\pm0.062$ mag, confirming the reliability of both methods. The bottom panel of Figure \ref{fig:3} compares the distances from the two methods, showing an average difference of $-16\pm191$ pc. We adopt the average of the distances from both methods as the final distances, resulting in a sample of 3,452 DCEPs, with an average relative distance error of 3.1\%. 

It should be noted that we adopted a constant extinction law in this work. However, as extinction laws can vary, the distance errors might be underestimated. This is also why we use infrared bands to minimize the impact of extinction uncertainties. The 3D spatial distribution of these DCEPs is presented in Figure \ref{fig:4}, and their parameters, including distances and extinctions, are listed in Table \ref{tab:2}.

\subsection{Comparison with \textit{Gaia} PWZ Distances}\label{compare gaia}
\begin{figure}[htb]
\figurenum{5}
\gridline{\fig{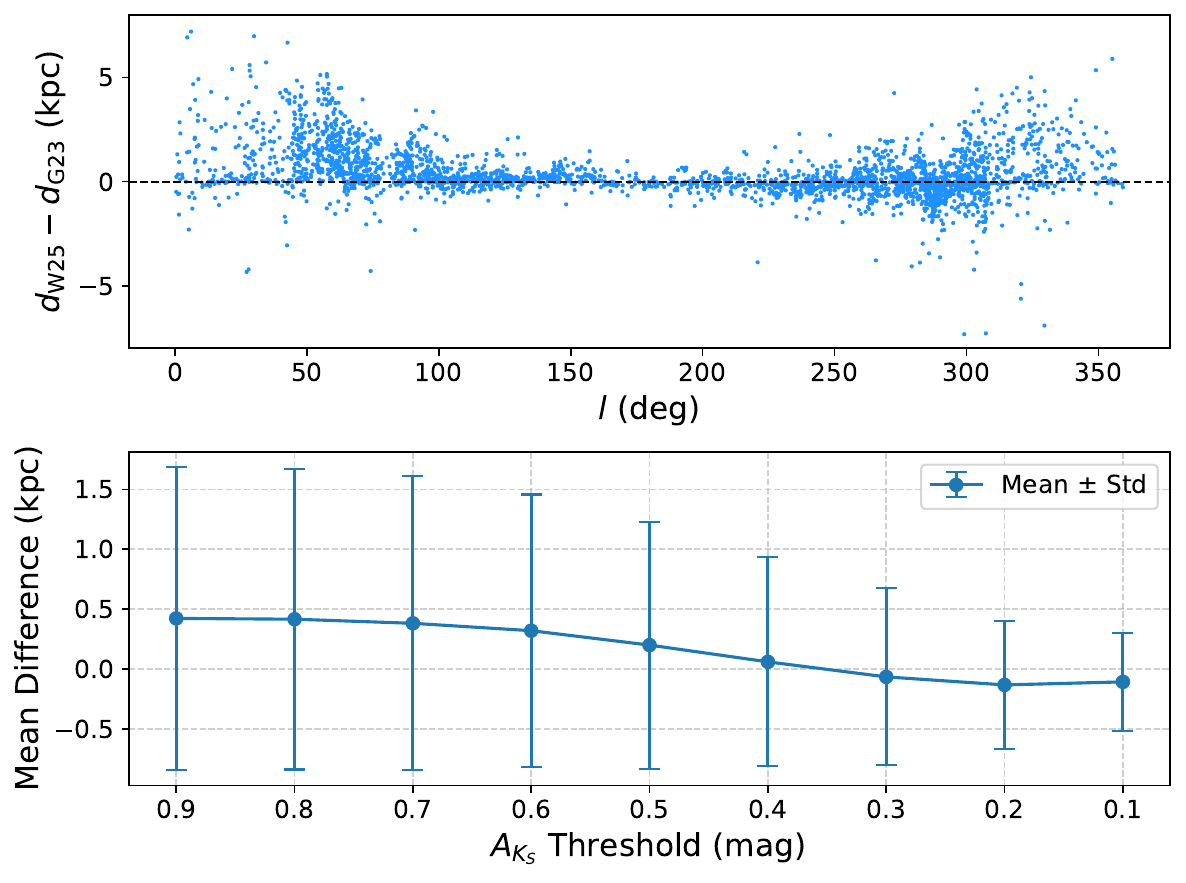}{0.45\textwidth}{}}
\caption{The top panel shows a comparison between our derived distances and those reported by G23, based on a sample of 2,953 DCEPs. The average difference is $422\pm1262$ pc. Meanwhile, the bottom panel shows the average difference obtained after imposing an upper limit on the extinction of the sample.
\label{fig:5}}
\end{figure}

The top panel of Figure \ref{fig:5} compares our distances with those from G23, revealing significant discrepancies toward the Galactic center. We find that the unreliability of $m_{G_{\text{BP}}}$ at the faint end is not the primary cause of this discrepancy, as excluding DCEPs with $m_{G_{\text{BP}}} > 20.3$ mag does not significantly reduce the difference. 
The main reason lies in the adoption of a constant optical extinction law (i.e., $\lambda=1.9$) to define the Wesenheit magnitude in the \textit{Gaia} bands: $w_G = m_G - \lambda \times(m_{G_{\text{BP}}} - m_{G_{\text{RP}}})$. However, the optical extinction law varies significantly across the MW disc \citep{wang17}, and the color term $m_{G_{\text{BP}}} - m_{G_{\text{RP}}}$ can reach values as high as 7 mag. This explains why the discrepancies are more pronounced toward the Galactic center with higher extinction. The bottom panel of Figure \ref{fig:5} illustrates the average difference obtained when we impose an upper limit on the extinction of the sample. It shows that as extinction decreases, the discrepancy between the two methods gradually diminishes. This indicates that the Wesenheit method in optical bands is only reliable in low-extinction regions. Using our derived distances as a reference and applying the PWZ relation in the \textit{Gaia} bands, we derive an average value of $\lambda=1.83\pm0.13$.

\begin{figure}[htb]
\figurenum{6}
\gridline{\fig{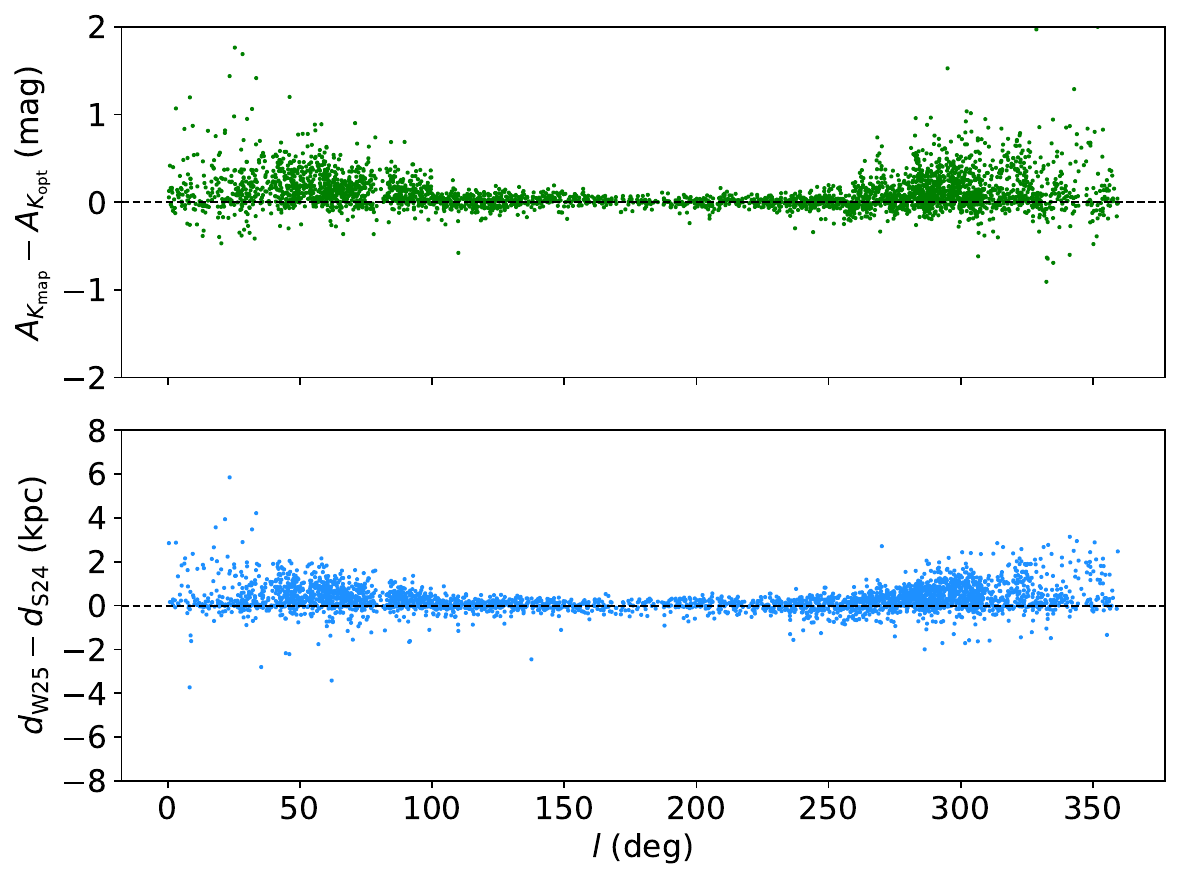}{0.45\textwidth}{}}
\caption{The top panel presents a comparison between our derived extinctions and those derived from \textit{mwdust}. Meanwhile, the bottom panel illustrates a comparison between our distances and those reported by S24, based on a sample of 3,318 DCEPs. The average difference in distance is $313\pm621$ pc.
\label{fig:6}}
\end{figure}

\begin{splitdeluxetable*}{ccccccccccccBccccccccc}
%\rotate
\setlength{\tabcolsep}{0.7pt}
\tabletypesize{\scriptsize}
\tablewidth{0pt}
%\tabletypesize{\tiny}
\tablenum{2}
\tablecaption{Distances and extinctions for our 3,452 Galactic DCEPs \label{tab:2}}
\tablehead{
\multicolumn{3}{c}{} &\multicolumn{9}{c}{Optimal distance method} & \multicolumn{9}{c}{BP-RP method}\\
\cline{4-12} \cline{13-21}
\colhead{Cepheid} &\colhead{$\langle d \rangle$}&\colhead{$\langle A_{K_{S}} \rangle$} &\colhead{$d_J$}&\colhead{$d_H$} &\colhead{$d_{K_{S}}$} &\colhead{$d_{W1}$} &\colhead{$d_{W2}$} &\colhead{$d_{[3.6]}$} &\colhead{$d_{[4.5]}$} &\colhead{$d_{opt}$} &\colhead{$A_{K_{_{opt}}}$} &\colhead{$d_J$}&\colhead{$d_H$} &\colhead{$d_{K_{S}}$} &\colhead{$d_{W1}$} &\colhead{$d_{W2}$} &\colhead{$d_{[3.6]}$} &\colhead{$d_{[4.5]}$} &\colhead{$d_{BR}$} &\colhead{$A_{K_{BR}}$} \vspace{-2mm}\\
\colhead{}  &\colhead{(kpc)} &\colhead{(mag)} &\colhead{(kpc)} &\colhead{(kpc)} &\colhead{(kpc)} &\colhead{(kpc)} &\colhead{(kpc)} &\colhead{(kpc)} &\colhead{(kpc)} &\colhead{(kpc)} &\colhead{(mag)} &\colhead{(kpc)} &\colhead{(kpc)} &\colhead{(kpc)} &\colhead{(kpc)} &\colhead{(kpc)} &\colhead{(kpc)} &\colhead{(kpc)} &\colhead{(kpc)} &\colhead{(mag)}\\
}
\vspace{-2mm}
\colnumbers
\startdata
        AA Mon & 4.354(0.109) & 0.220  & 4.324(0.442) & 4.393(0.279) & 4.415(0.191) & 4.442(0.178) & 4.428(0.229) & 4.129(0.161) & 4.349(0.160) & 4.339(0.108) & 0.232  & 4.466(0.457) & 4.472(0.284) & 4.463(0.193) & 4.463(0.179) & 4.442(0.230) & 4.148(0.162) & 4.363(0.161) & 4.369(0.110) & 0.208  \\ 
        CS Mon & 4.004(0.121) & 0.143  & 4.001(0.412) & 4.015(0.258) & 4.020(0.179) & 3.734(0.152) & 4.062(0.212) & 4.094(0.153) & 4.095(0.148) & 3.995(0.125) & 0.150  & 4.083(0.420) & 4.060(0.261) & 4.047(0.180) & 3.746(0.152) & 4.070(0.213) & 4.106(0.154) & 4.103(0.148) & 4.013(0.118) & 0.135  \\ 
        CV Mon & 1.842(0.056) & 0.194  & 1.873(0.192) & 1.850(0.117) & 1.838(0.079) & 1.931(0.084) & 1.946(0.102) & 1.792(0.063) & 1.810(0.061) & 1.844(0.058) & 0.191  & 1.858(0.190) & 1.842(0.117) & 1.834(0.079) & 1.928(0.084) & 1.944(0.102) & 1.789(0.063) & 1.808(0.061) & 1.840(0.054) & 0.196  \\ 
        V872 Cen & 9.535(0.175) & 0.303  & 9.704(0.997) & 9.438(0.605) & 9.305(0.408) & 9.631(0.386) & 9.769(0.505) & 9.447(0.377) & 9.740(0.399) & 9.558(0.179) & 0.296  & 9.505(0.977) & 9.347(0.599) & 9.263(0.407) & 9.592(0.384) & 9.729(0.503) & 9.421(0.376) & 9.704(0.397) & 9.512(0.171) & 0.310  \\ 
        V414 Vul & 11.360(0.268) & 0.550  & 11.545(1.189) & 11.170(0.726) & 11.416(0.500) & 11.750(0.471) & 11.573(0.598) & 11.569(0.499) & 11.065(0.474) & 11.444(0.244) & 0.528  & 10.849(1.117) & 10.825(0.704) & 11.203(0.491) & 11.600(0.465) & 11.475(0.593) & 11.445(0.494) & 11.004(0.472) & 11.275(0.292) & 0.572 \\  
\hline
\enddata
\tablecomments{The table includes the following information: (1) DCEP names; (2) and (3) the average distances and extinctions derived from the optimal distance and BP-RP methods, respectively; (4)–(12) parameters derived using the optimal distance method; and (13)–(21) parameters derived using the BP-RP method. The complete table is available at CDS.}
\end{splitdeluxetable*}

\subsection{Comparison with 3D Extinction Map Distances}\label{compare map}
 
S24 derived DCEPs distances using the $W1$ band PL relation \citep{wang18} and the 3D extinction map \textit{mwdust} from \cite{bovy16}. The top panel of Figure \ref{fig:6} compares our extinctions with those from \textit{mwdust}. It shows that in the low-extinction region toward the anti-Galactic center ($100^{\circ} < l < 250^{\circ}$), the two extinction values agree well. However, in the high-extinction region toward the Galactic center, significant discrepancies arise, with extinction values from \textit{mwdust} being noticeably higher than ours. Consequently, as shown in the bottom panel of Figure \ref{fig:6}, distances derived by S24 in this region are systematically smaller than ours. We test different extinction laws, but they are insufficient to explain the significant extinction differences observed in the Galactic center region.

The maximum extinction $A_K$ we derive is approximately 1.4 mag, whereas the maximum $A_K$ derived from \textit{mwdust} can reach values as high as 2.5 mag. This discrepancy is significant, as it directly impacts the derived distances and the detectability of DCEPs. For example, assuming the brightest DCEP ($\log P_{\textrm{F}} \approx 2$) and the \textit{Gaia} detection limit of $m_G \approx 20.7$ mag, with $A_K=1.4$ mag, and using the $G$ band PLZ relation from T24a, the calculated distance is approximately 4.5 kpc. However, DCEPs with $A_K>1.4$ mag, as derived from \textit{mwdust}, are located beyond this distance, exceeding \textit{Gaia}'s detection capabilities. Additionally, it should be noted that \textit{mwdust} does not cover the entire MW and is generally considered reliable only within 5 kpc from us. This further highlights the limitations of applying \textit{mwdust} to DCEPs at greater distances.

\section{Conclusions}\label{sec:conclusions}
We calibrate the PLZ relations in three NIR bands ($J$, $H$, $K_S$) and four MIR bands ($W1$, $W2$, $[3.6]$, and $[4.5]$) using the parallaxes of the OC-DCEPs and the field DCEPs from \textit{Gaia} DR3. Notably, the PLZ relations for the $W1$ and $W2$ bands are calibrated for the first time. The metallicity coefficients we derived range from -0.30 to -0.46.
Using our calibrated PLZ relations, we estimate $\mu_{\textrm{LMC}}$, which shows excellent agreement with the geometric distance modulus determined by \cite{pie19}, thereby validating the reliability of our calibration. Applying the PLZ relations calibrated in this work, together with the optimal distance method and the BP-RP method, we derived distances and extinctions for 3,452 Galactic DCEPs. All of these DCEPs have relative distance errors below 10\%, with an average relative error of 3.1\%. This study provides a robust dataset for future investigations into the structure and evolution of the MW.

%\begin{acknowledgments}
%This work was funded by the NSFC Grands 11933011, National SKA Program of %China (Grant No. 2022SKA0120103) and the
%Key Laboratory for Radio Astronomy. L.Y.J. thanks the support of the NSFC %grant No. 12203104, the Natural Science Foundation of Jiangsu Province
%(grant No. BK20210999), the Entrepreneurship and Innovation Program of %Jiangsu Province.
%\end{acknowledgments}

\bibliography{ApJ}{}
\bibliographystyle{aasjournal}


\end{CJK}
\end{document}